\documentclass[11pt,twoside]{article}

\usepackage{asp2006}
\usepackage{epsf}
\usepackage{lscape}

\begin{document}
\title{Barred Galaxies in the Coma Cluster}   %%% Fill in title
\author{Irina Marinova,\altaffilmark{1} 
Shardha Jogee,\altaffilmark{1} 
Neil Trentham,\altaffilmark{2} 
Henry C.~Ferguson,\altaffilmark{3}
Tim Weinzirl, \altaffilmark{1}
%Bahram Mobasher,\altaffilmark{3} 
Marc Balcells,\altaffilmark{4}
David Carter,\altaffilmark{5}
Mark den Brok, \altaffilmark{6}
Peter Erwin,\altaffilmark{7} 
Alister W.~Graham,\altaffilmark{8} 
Paul Goudfrooij, \altaffilmark{3}
Rafael Guzm\'an, \altaffilmark{9}
Derek Hammer, \altaffilmark{10}
Carlos Hoyos, \altaffilmark{9}
Reynier F.~Peletier, \altaffilmark{6}
Eric Peng, \altaffilmark{11,12}
and Gijs Verdoes Kleijn, \altaffilmark{6}
 }   %%% Fill in author names
\altaffiltext{1}{Department of Astronomy, University of Texas at
Austin, Austin, TX}   %%% Fill in author affiliations
\altaffiltext{2}{Institute of Astronomy, Madingley Road, Cambridge CB3 0HA}
\altaffiltext{3}{Space Telescope Science Institute, 3700 San Martin Drive, 
Baltimore, MD 21218, USA.}
\altaffiltext{4}{Instituto de Astrof\'isica de Canarias, 
38200 La Laguna, Tenerife, Spain.}
\altaffiltext{5}{Astrophysics Research Institute, Liverpool John Moores 
University,  Birkenhead, UK.}
\altaffiltext{6}{Kapteyn Astronomical Institute, University of Groningen, Groningen, The Netherlands.}
\altaffiltext{7}{Max-Planck-Institut f\"ur Extraterrestrische Physik, Garching, Germany.}
\altaffiltext{8}{Centre for Astrophysics and Supercomputing, Swinburne 
University, Hawthorn, Australia.}
\altaffiltext{9}{Department of Astronomy, University of Florida, Gainesville, FL 32611, USA.}
\altaffiltext{10}{Department of Physics and Astronomy, Johns Hopkins, Baltimore, MD 21218, USA.}
\altaffiltext{11}{Department of Astronomy, Peking University, Beijing 100871, China}
\altaffiltext{12}{Kavli Institute for Astronomy and Astrophysics, Peking University, Beijing 100871, China}

\begin{abstract} %%% Abstract to run on from here.
We use ACS data from the $HST$ Treasury survey of the Coma cluster
($z\sim$~0.02) to study the properties of barred galaxies in the Coma core, 
the densest environment in the nearby Universe. This study provides a 
complementary data point for studies of barred galaxies 
as a function of redshift and environment. From $\sim$~470 cluster 
members brighter than $M_{\rm I} = -11$~mag, we select a sample of 
%%%%%%%
46 disk galaxies (S0--Im) based on visual
classification.  The sample is dominated by S0s for which we find
an  optical bar fraction of 47$\pm$11\% through ellipse fitting 
and visual inspection. 
Among the bars in the core of the Coma cluster, we do not find any very large 
($a_{\rm bar} > 2$~kpc) bars. 
%% or many very strong ($e_{\rm bar} >$ 0.7) bars.
Comparison to other studies reveals that while the optical bar fraction 
for S0s shows only a modest variation across 
low-to-intermediate density environments (field to 
intermediate-density clusters),
it can be higher by up to a factor of $\sim$~2 in the very 
high-density environment of the rich Coma cluster core.
\end{abstract}

\section{Introduction}   

Bars are the most efficient internal driver of secular evolution in
disk galaxies. They efficiently redistribute angular momentum in the disk 
and drive gas to the central regions of galaxies 
where it can pile up and initiate powerful starbursts \citep{schwarz81, 
 sakamoto99, jogee05}. In this way, bars are thought to
build disky
central components known as pseudobulges
\citep{kormendy79, combes90, kormendy93}.

How does the frequency of barred disk galaxies vary
across different environments and what does this imply about
the evolution of bars and their host galaxies? Quantitative results
addressing this issue are only starting to emerge. Several recent studies (\citealp{barazza09,
aguerri09, marinova09}, hereafter M09) find that the optical bar fraction shows at most a modest variation 
($\pm10\%$) between field and intermediate density environments (e.g., groups 
and moderately rich clusters). However, some studies \citep{barazza09, thompson81,
  andersen96} have suggested that, within a galaxy cluster, the
bar fraction is higher in the dense core regions than the outskirts. 
This remains an open question due to issues such as limited number statistics in the
core and  uncertainties in cluster membership. 

We 
explore these questions in Coma, the richest cluster
in the nearby Universe. Our results provide a comparison point for 
studies of barred galaxies in field and group environments in the
nearby universe and at high redshift.

\section{Data and Cluster Sample}
Our data come from the \textit{Hubble Space Telescope} ($HST$)
Advanced Camera for Surveys (ACS) Treasury 
survey of the Coma cluster at $z\sim$~0.02 \citep{carter08}. Due to
the 2007 failure of ACS, the survey is only $\sim$~28\% complete.
The available data predominantly cover
the cluster core and map
an area of 274 arcmin$^{2}$ down to a limiting
magnitude of $I =$~26.8 mag in F814W (AB mag). Source extraction
yields 472 galaxies with $I\le$~24~mag ($M_{\rm I} \sim -11$~mag). 
Cluster members are picked from the membership catalog of 
Trentham et al.~(in preparation) where galaxies are assigned a 
membership class $C$ from 0 to 4.
Galaxies with $C=0$ are spectroscopically confirmed members.
Galaxies without  spectroscopic redshifts are assigned 
a membership likelihood based on surface brightness and
morphology, such that $C$~=~1 to 3 represent very likely to 
plausible members. 
Out of the 472 galaxies with $I\le$~24~mag, 470 are considered as 
cluster members, with the breakdown in classes 0 to 3 being 
41\%, 7\%, 27\%, 25\%. We use this sample for the rest of the study.  

All the 470 galaxies were assigned visual morphological types 
(Trentham et al. in  prep.), which include 
ellipticals (E), lenticulars (S0), spirals (Sa--Sm), irregulars 
(Im),  and dwarfs. Where available, Hubble types were taken from
the Third Reference Catalog of Bright Galaxies (RC3;
\citealp{deVauc1991}), otherwise galaxies were classified into Hubble
types based on a visual estimate of the prominence of the bulge, 
tightness of the spiral arms, and clumpiness of the disk. 
% NEED TO DESCRIBE MORPHOLOGICAL CLASSIFICATION BEFORE THE 
% SENTENCE BELOW  SINCE YOU TALK OF DWARFS HERE 
The absolute magnitude distribution for the total cluster sample, and the
breakdown for different morphologies is shown in Fig~1a. 
Although we use visual classification without further magnitude cuts to 
identify dwarfs, it is good to note that the visually identified dwarfs 
are faint as expected, and have $M_{\rm I}\ge -17$~mag (Fig.~1a).

As bars are intrinsically a disk phenomenon, it is the norm to 
characterize the frequency of large-scale or primary 
bars by citing the fraction $f_{\rm bar-opt}$  
of \textit{disk galaxies} with bars,  where disk galaxies 
are defined to be systems harboring a large-scale outer disk.
At the bright end, disk galaxies include S0s and Sa 
to Im, while among dwarf systems, they include dwarf irregulars
(dIrr) and potentially some dS0s with disk components. 
In these proceedings, our goal is to measure  $f_{\rm bar-opt}$  
for non-dwarf disk galaxies of Hubble types S0 to Im, 
while the  investigation 
of bars in dwarf galaxies will be described in a future paper.
The first challenge is to identify  the disk galaxy sample (S0s 
to Im). 
The danger in identifying disks only by color or S{\'e}rsic cut in 
clusters has been discussed extensively by M09, and we therefore 
use the visual classifications for the Coma sample described above. 
We select all disk galaxies ranging from  S0 to
Im, excluding two galaxies that are
partially off the edge of a tile and one that is a distorted merger
remnant. This brings our final disk sample to
%%%%%%%%%%%%
46 galaxies.
%%%%%%%%%%%%% 
As illustrated in Fig.~1b,  the disk sample is dominated by S0s 
as expected from the morphology-density relation.

\section{Bar Identification}
We use ellipse fitting to identify bars and also perform an extra
check through visual classification for all galaxies. 
As described in our earlier work 
(\citealp{jogee04, mj07}, hereafter MJ07; M09), 
we perform ellipse fitting and then classify galaxies as 
inclined, barred, or unbarred, based on the
radial profiles of ellipticity ($e$) and position angle (PA), as well 
as overlays of the fitted ellipses onto the galaxy images.  
A galaxy is classified as `inclined' if the outer disk inclination $i \ge
60^{\circ}$. Following common practice, we exclude the 16 inclined galaxies 
that we find because it is difficult to identify morphological features 
in highly inclined galaxies. 
From the remaining sample of moderately inclined disks, we
classify a galaxy as barred if (1)~the $e$ rises to a global maximum,
$e_{\rm bar} > 0.25$, while the PA remains relatively constant (within 
$20^{\circ}$), 
and (2)~the $e$ drops by at least 0.1 and the PA 
changes at the transition between the bar and disk region. The change
in PA between the bar and disk region is statistically expected, as the line of nodes for
the bar and disk are oriented independently. 

These characteristics work well in lower-density environments
(see MJ07), where disk galaxies with extremely large
bulge-to-disk ratios are rare.  However in the core of the Coma cluster
such galaxies exist, and when the bar is oriented perpendicular to
the disk major axis, the ellipticity of the bar isophotes can be
diluted by the extremely large bulge.
As a result, the above criterion (1) is not satisfied: 
specifically, the peak bar $e$ in the ellipticity
radial profile is not the
global maximum, and the PA presents a `bump' at the location of the
bar instead of remaining flat. 
There are three such cases in our sample (marked `vis' in Fig.~1c), 
and for these we use visual
inspection to classify the galaxy as barred. We note that in field and
intermediate-density samples, the bar signature may sometimes be masked in the
optical by dust and patchy star formation. However, in the dense Coma
core, we expect dust extinction to have little effect. 

Overall in the 
sample of moderately inclined disks of type S0--Sm, 
%%%%%%%%%%%%%
we find 10 barred and 20 unbarred systems through
ellipse fitting  and visual classification.
%%%%%%%%%%%%%
 Our classifications 
are summarized in Table~1 and barred galaxies are shown in 
Fig.~1c.

\section{Results and Comparison to Studies Across Different Environments}
The  optical bar fraction  $f_{\rm bar-opt}$  across the disk sample
of S0 to Im galaxies is given by  $N_{\rm bar}$/($N_{\rm bar} + 
N_{\rm unbar}$), where $N_{\rm bar}$ and $N_{\rm unbar}$ represent 
the number of barred and unbarred disk galaxies, respectively,  in the 
moderately inclined sample. 
%%%%%%%%%%%%
We find  that $f_{\rm bar-opt}$, 
averaged across (S0-Im) galaxies, is 10/30 or 33$\pm$9\%  (Table~1).
%%%%%%%%%%%%%%%%%

However, compiling an average bar fraction across a wide range in 
Hubble types  only gives limited insight since recent studies show that
the optical bar fraction is a strong function of host galaxy properties,
such as the bulge-to-disk ratio ($B/D$) and  galaxy luminosity.
Specifically, \citet{barazza08} and M09 show 
that the optical bar fraction at $z < 0.03$ is highest in
galaxies that are disk-dominated and have very low $B/D$. In addition,
in M09 we find in the A901/902 clusters at $z\sim$~0.165 that for a
given morphological class, $f_{\rm bar-opt}$  is higher for brighter galaxies.
Unfortunately, the Coma sample is too small to be split into 
bins of $B/D$, luminosity, and Hubble types. However, it is clear
from Table~1  and Fig.~1b that the sample of disks is dominated by 
S0s and that the optical bar fraction is driven by these galaxies.
Therefore, when comparing  our results on Coma to other studies, 
we focus on the optical bar fraction for S0s.
We find  $f_{\rm bar-opt}$ for S0s is 47$\pm$11\% (row 2 of Table 1).
We consider this value to be an upper limit because it is likely 
that we are missing some unbarred S0s, which are easily confused with
ellipticals.
%%%%%%%%%%%%

In Table~2, we show a
comparison with: a previous study of the Coma cluster by 
Thompson (1981, T81) using
visual classification to identify bars; M09 for barred
disks using ellipse fitting in the Abell 901/902 cluster system at
$z\sim$~0.165; bars in Virgo (Giordano et al.~2010, in
prep., G10); and Aguerri et
al.~(2009, A09) who study barred
disks using ellipse fitting and Fourier decomposition at $z\sim$~0.01--0.04 in
environments ranging from the field to intermediate densities 
comparable to cluster outskirts. 

%%%%%%%%%%%%%%%%%%%%
\begin{table}[!ht]
\caption{Disk sample classifications based on ellipse fits and visual
  inspection ($N_{\rm total} = $~46).}
\smallskip
\begin{center}
{\small
\begin{tabular}{cccccc}
\tableline
\noalign{\smallskip}
   & All & Highly inclined & Unbarred & Barred & $f_{\rm bar,opt}$\\
\noalign{\smallskip}
\tableline
\noalign{\smallskip}
S0--Sm & 46 & 16 & 20 & 10 & 33$\pm$9\%\\
S0     & 30 & 13 & 9 & 8  & 47$\pm$11\%\\
S0--Sab& 35 & 14 & 12 & 9  & 43$\pm$11\%\\
Sb--Sm & 11 & 2  & 8 & 1  & 11$\pm$10\%\\ 
\noalign{\smallskip} 
\tableline
\end{tabular}}
\end{center}
\end{table}
%%%%%%%%%%%%%%%%%%%%

%%%%%%%%%%%%%%%%%%%%
\begin{table}[!ht]
\caption{Optical bar fraction $f_{\rm bar-opt}$ for
  different environments.}
\smallskip
\begin{center}
{\small
\begin{tabular}{cccc}
\tableline
\noalign{\smallskip}
Study & Environment & S0 \\
\noalign{\smallskip}
\tableline
\noalign{\smallskip}
this work & Coma core, $z\sim$~0.02 & 47$\pm$11\% (upper limit) \\
T81  & Coma core, $z\sim$~0.02 & 34$\pm$6\%$^{a}$  \\
T81  & Coma core, $z\sim$~0.02 & 42$\pm$7\%$^{b}$ \\
M09  & A901/902 clusters, $z\sim$~0.165 & 25$^{+10\%}_{-3\%}$ \\
G10  & Virgo, $z\sim$~0 & 36$\pm$9\%  \\
A09  & field--intermediate, $z\sim$~0.01--0.04 & 29\%  \\
\noalign{\smallskip} 
\tableline
\end{tabular}}
\end{center}
\small{\textit{a}: from raw galaxy counts}\\
\small{\textit{b}: after correcting galaxy counts for projection effects}
\end{table}
%%%%%%%%%%%%%%%%%%%%
  
Table~2 shows that for the core of the Coma cluster, our 
optical bar fraction for S0 
%%%%%%%%%%%%%%%%%%%%
galaxies (47$\pm$11\%) is consistent with the projection-corrected
bar fraction (42$\pm$7\%) from T81.
%%%% higher than that found by T81 (34$\pm$6\%) from 
%%%%  raw galaxy counts in the Coma core. 
%%%% Additionally, T81 calculates a bar fraction for S0s of 42$\pm$7\% after
%%%% galaxy counts in the Coma core are corrected for projection effects. 
% This is consisent 
% with our results. 
T81 also finds that the (corrected) optical bar fraction for S0s decreases from
42$\pm$7\% in the Coma core to 22$\pm$3\% in the outskirts and
outer regions of Coma. We cannot test the latter result directly 
as we do not have  ACS data in the Coma outskirts. 
However, we see from  Table~2 that while the optical 
bar fraction for S0s shows only a modest variation across 
low-to-intermediate density environments (field to 
intermediate-density clusters; last 3 rows of Table 1),
it can be higher by up to a factor of $\sim$~2 in the very 
high-density environment of the rich Coma cluster core.

%%%%%%%%%%%%%%%%%%%%
%shows only a modest variation  (by a 
%factor of less than two) across a very wide range in environmental 
%densities, spanning the dense core of the
%rich Coma cluster,  the intermediate-density Abell 901/902 clusters, 
%the low density Virgo cluster, and field regions.

\begin{figure}[!ht]
\vspace {-0.5 in} 
\plotone{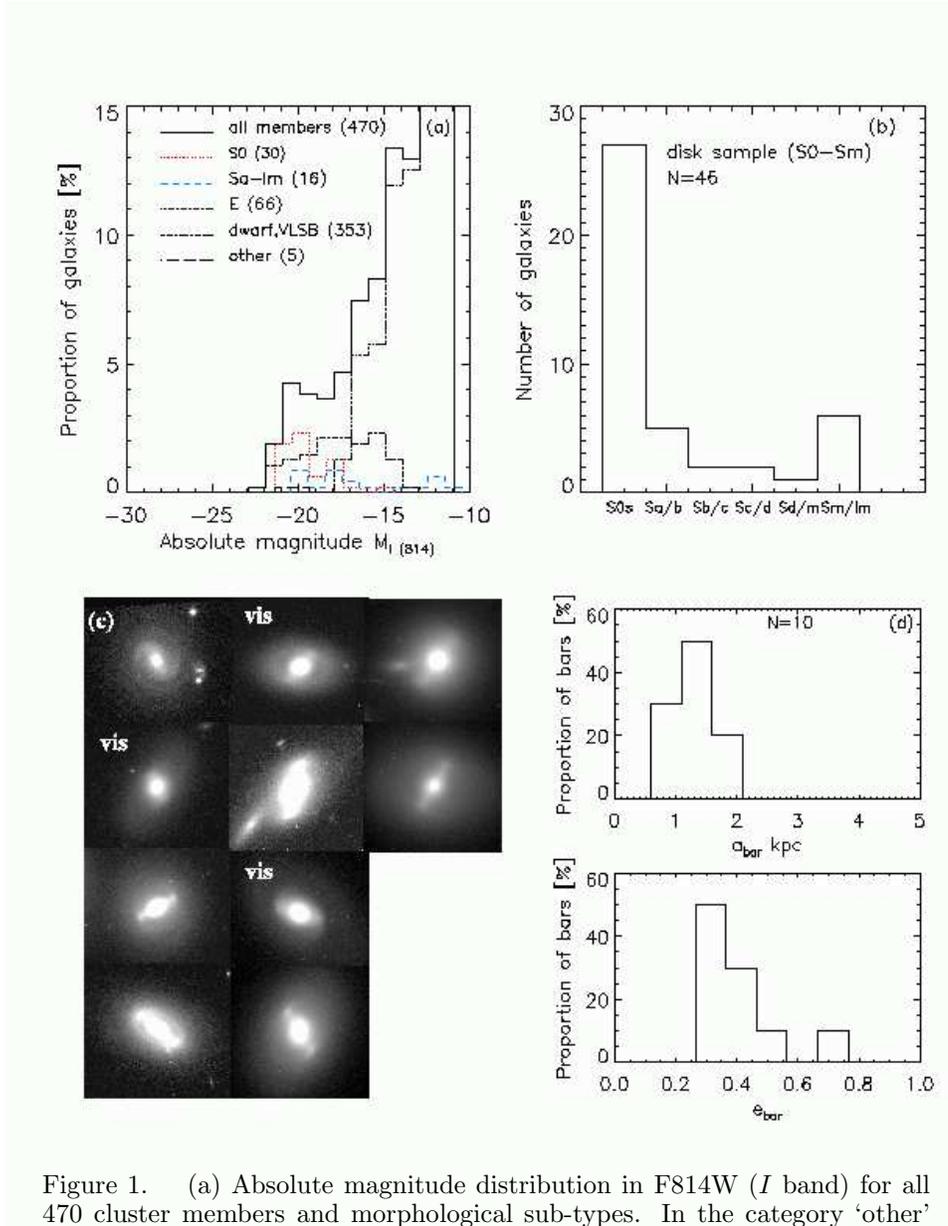}
\vspace {-0.5 in} 
\caption{(a)~Absolute magnitude distribution in F814W ($I$ band) for
  all 470 cluster members and morphological
  sub-types. In the category `other' we include two galaxies
  classified as `peculiar', two that are partially off the edge of a
  tile, and one distorted merger remnant.  
(b)~Morphology distribution of our sample of disk
  galaxies. Most of the disks are S0s.  
(c)~Barred galaxies identified through ellipse fitting and visual
 inspection (marked `vis').
(d)~Distributions of $a_{\rm bar}$ (top) and $e_{\rm bar}$ (bottom).
% and the Abell 901/902  cluster (M09). 
}
\end{figure}

Finally, in Fig.~1d we show the semi-major axis $a_{\rm bar}$ 
and the strength (ellipticity) $e_{\rm bar}$  of the 10 bars 
we identified in the core of Coma. 
While the number statistics are very small, it is interesting
that we do not find any large bars ($a_{\rm bar} >$~2 kpc), 
as is seen in S0s in less dense environments. 
This may reflect evolutionary processes among S0s and bars in the
dense environment of the Coma cluster core.

%\acknowledgements %%% Text of acknowledgements runs on after this command.

%%% THE BIBLIOGRAPHY
%%%
%%% CONSULT SECTION 3 OF "INSTRUCTIONS FOR AUTHORS" FOR HOW TO USE NATBIB.
%%% AUTHORS ARE ENCOURAGED TO USE EITHER THE "THEBIBLIOGRAPY" ENVIRONMENT
%%% BY UNCOMMENTING (DELETING THE "%" SYMBOL) THE COMMANDS BELOW, OR BY
%%% USING THE BIBTEX ENVIRONMENT. TO FIND OUT WHICH IS APPLICABLE TO YOUR
%%% CONTRIBUTION, CONSULT THE VOLUME EDITORS FOR YOUR PROCEEDINGS.
%%%

\end{document}